\documentstyle[aps,prl,multicol,epsf]{revtex} 

\begin{document}

\title{
Nonuniversal scaling behavior of Barkhausen noise }

\author{Bosiljka Tadi\'c$^\star $}

\address{
Jo\v{z}ef Stefan Institute, University of Ljubljana, 
P.O. Box 3000, 1001-Ljubljana, Slovenia }


\maketitle
\begin{abstract}
\newline
We simulate  Barkhausen avalanches on  fractal clusters in a 
two-dimensional diluted Ising ferromagnet with an effective Gaussian 
random field. We vary the concentration of defect sites $c$ 
and find a scaling region for moderate disorder, where the distribution 
of avalanche sizes has the form $D(s,c,L) = 
s^{-(1+\tau (c))}{\cal{D}}(sL^{-D_s(c)})$.  The  exponents $\tau (c)$ 
for size and $\alpha (c)$ for length distribution, and the fractal 
dimension of avalanches $D_s(c)$ satisfy the scaling relation 
$D_s(c)\tau (c) =\alpha (c)$. For fixed disorder  the exponents  vary  
with driving rate in  agreement with  experiments on amorphous  
Si-Fe alloys. 

\end{abstract}

\begin{multicols}{2}
Barkhausen noise (BN) occurs at low temperatures when a {\it disordered }
ferromagnetic sample is slowly driven by the external magnetic field.
A small ramp in the field triggers one domain, and  the perturbation 
spreads to the neighboring domains  producing an avalanche, which manifests 
itself as a series of jumps in the magnetization as the system passes from 
one metastable state to another. Hysteresis in the field-magnetization 
plane is inherent to these processes. If the field is ramped up and down 
close to the coercive field of the sample, the avalanches show a broad 
distribution of sizes, although the triggering mechanisms are strictly  
local and fully deterministic. The avalanche-like dynamics is controlled 
by the spatial distribution of quenched-in defects of various types, and is
not influenced by the thermal fluctuations (as long as the system is
below its ordering temperature). 

Besides its practical applications \cite{Sipahi}, Barkhausen noise has been
studied both experimentally \cite{ex1,ex2,ex3Bgg,ex4,ex5,ex6,ex7Durin} and 
theoretically \cite{ex3Bgg,num1,num2,num3,num4,num5} as an interesting 
example of collective dynamical behavior far from equilibrium. 
Recent experimental studies of BN were made on amorphous
alloys such as metglass 2605S-2 and alumel \cite{ex1}, metglass 
2606TCA \cite{ex2}, VITROVAC 6025X \cite{ex3Bgg}, and on  
Fe-Ni-Co  \cite{ex4,ex5} and Si-Fe  \cite{ex5,ex6,ex7Durin} alloys. 
Numerous data collected in these
and other experiments show that the power spectrum decays with frequency
as $\sim \omega ^{-\phi }$ (1.5$\le \phi \le $2) and the distributions of 
size, duration, and energy
associated with Barkhausen jumps exhibit a power-law behavior over a few
decades with a cutoff. More detailed analysis reveals \cite{ex3Bgg} that
the measured critical exponents for various distributions obey certain 
scaling relations, which are also derived in Ref.\ \cite{ex3Bgg} assuming 
a  definite type of the elementary pulse. 

Experiments were typically done on one sample and no special care was 
taken in controlling the strength (and  type) of disorder,
which is responsible for avalanche-like dynamics. Other details
which influence  the statistics of BN were studied. In particular,
it was found in Ref.\ \cite{ex7Durin} that in 1.8\% Si-Fe alloy
the exponents characterizing the distributions of size and duration of 
Barkhausen pulses appear to vary linearly with the driving rate of the 
external field.  Urbach {\it et al.} \cite{ex4} studied  the importance of 
demagnetizing effects.

The dynamics of field-driven random ferromagnets at zero temperature
has been studied by  numerical simulations using  Ising model with
 random fields (RFIM) \cite{num1,num2} and with random bonds of the 
 spin-glass type (RBIM) \cite{num4,num5}. It has been recognized 
that in both cases a critical level of disorder exists  at which 
the  response to the external field becomes slow, with a broad  
distribution of avalanche sizes. This disorder-driven phase transition 
has been reported to have unusually large critical region---50\% above 
the critical point---in the case of 3-dimensional RFIM \cite{num2}.

In the present work we would like to point out the relevance of another 
type of defects to the scaling behavior of BN. 
In magnetic systems random fields are generated on a semimacroscopic 
scale  by other types of disorder (i.e., random site or random bond defects)
when the system is placed in an external magnetic field \cite{ImryMa}. 
At the asymptotic thermally driven phase transition random fields dominate, 
leading to a new universality class of (equilibrium) phase transitions 
\cite{RGRFM}. 
However, far from equilibrium  where the  Barkhausen noise is 
produced, the impurities of this  type  may  still be relevant and are
competing with the random fields in  the pinning of domain boundaries. 
In particular, in the case of Si-Fe alloys the sample  is diluted by 
nonmagnetic ions, whereas in the case of Fe-Ni-Co alloys and  metglass 
samples  the type of defects can be  described by the random 
anisotropy and random-bond models, respectively.  

We consider the diluted ferromagnetic Ising model with concentration 
$c$ of empty (i.e., zero spin) sites, and add a small Gaussian random 
field with variance  $f$  at each occupied site. 
It should be stressed that our model differs from the ones
considered  in Refs.\ \cite{num1,num2,num3,num4,num5} in two crucial 
aspects:  
(a) There is a larger parameter space in the $(c,f)$-plane,
 where the scaling behavior of the avalanches could  emerge, and  a
 well defined percolation threshold along the $c$-axis; (b)
For  $c\neq 0$ the avalanches contributing to  Barkhausen noise 
 develop on {\it noncompact clusters} of spins (see Fig.\ 1), 
the compactness of which is tuned  via the parameter $c$.
This leads to new prominent features of BN in diluted ferromagnets
\cite{dilmg}.

We start with a (coarse-grained) Ising model 
\begin{equation}
{\cal{H}} = -\sum _{i,j}J_{ij}S_iS_j - \sum _i(h_i+H)S_i  \equiv 
-\sum _ih_i^{loc}S_i \; \label{Hamiltonian}
\end{equation}
\noindent
where $S_i=\pm 1$ and the interactions are $J_{ij}=J_0$ between nearst 
neighbor occupied sites, and $J_{ij}=0$ if at least one site is a defect. 
We use a 2-dimensional square lattice with a fraction $c$ of randomly
distributed defect (i.e., $S_i=0$) sites and periodic boundary conditions.
As already mentioned, the local random fields $h_i$ are   generated on
a semimacroscopic scale. 
We assume Gaussian distributed random fields with zero mean and 
variance $f$ at each occupied site. The disorder is considered as quenched.

The system is driven by small ramps of the external field $\Delta H$
starting from a strong negative field $-H_{max}$ (for  practical 
reasons $H_{max}\le 4$). Parallel updating of spin states is applied 
according to the rule 
$S_i(t+1) = {\mathrm{sgn}} \ h_i^{loc}(t)$. 
For each rump of the external field the system is updated as long as 
there are no more unaligned spins. 
Clusters of flipped spins are monitored (an example is shown  in Fig.\ 1
\cite{U-th}) and the distribution of cluster sizes $D(s)$ is determined. 
The distribution is  integrated over  the  whole hysteresis loop
and averaged over total number of samples, ranging up to 400. 
We first show the results for a fixed driving rate $\delta h 
\equiv \Delta H /H_{max}$ = 0.01. 

In Fig.\ 2
the integrated probability distribution $D(s)$ of avalanches of
size $s$ or larger is shown  for fixed random-field variance $f=0.5$ and
for few  values of concentration $c$ (varied along the dashed line on 
the phase diagram  in Fig.\ 3). For very weak disorder close 
to the zero-temperature ferromagnetic fixed point (lower right corner 
in Fig.\ 3) the system is ordered and the external field flips 
the magnetization via a first order phase transition. Consequently, the
hysteresis loop is rectangular  and the integrated 
probability distribution shows zero slope (top curve in Fig.\ 2). 
At fixed $f=0.5$ a finite slope appears  first  for $c\approx 0.05$, 
increasing  gradually with $c$ in the range $0.05\mathopen< 
c \mathopen< 0.35$.
Assuming 
\begin{equation}
D(s)\sim s^{-\tau (c)} \;  \label{soc_s}
\end{equation}
\noindent
and fitting  the straight sections of the lines in Fig.\ 2 to Eq.\ 
(\ref{soc_s}), we  determine the exponent   $\tau (c)$ for 
$0.05\le c \le 0.3$. We find that the exponent \cite{thexp} $1+\tau (c)$ 
varies from 1.155$\pm$ 0.002 
at $c=$0.05 to 1.373 $\pm $  0.001 at $c=$0.3 (see inset to Fig.\ 1).
It should be noted that the loss of the power-law behavior at a
finite cutoff size is not exponential, but rather a  stretched exponential
with a $c$-dependent exponent (see also Ref.\ \cite{ex3Bgg}).
 
In order to study the cutoff systematically, we varied the size  
of the lattice $L$ at several  fixed values of $c$. 
For the integrated probability
distribution $D(s,c,L)$ the following scaling form is appropriate
\cite{TR}
\begin{equation}
D(s,c,L) = \ell ^{-\alpha }D(s\ell ^{-D_s},c\ell ^{\lambda _c},
\ell ^{-1}L)\; , \label{scal-gen}
\end{equation}
\noindent
where  $\alpha $ is the scaling exponent for the probability distribution
of avalanches of length $\ell $ or longer, and $D_s$ is the
fractal dimension of  avalanches (not to be confused with the fractal
dimension of percolation clusters). The exponent $\lambda _c$ is associated 
with the presence of defects. 
If the scaling function on the right hand side of  Eq.\ (\ref{scal-gen}) 
depends explicitly on $c\ell ^{\lambda _c}$ with $\lambda _c \ge 0$,   
the criticality (and the associated power-law behavior) could be 
maintained only if the disorder is kept at a critical value, say, 
$c_0$, i.e., we are dealing with a (nonequilibrium) phase transition. 
For $c\neq c_0$ the system is subcritical and the   coherence 
length  should be specified  as a  function of disorder $\xi (c-c_0)$. 
On the other hand, the parameter $c$ may  appear implicitly via  
tuning of the exponents $\alpha $ and $D_s$ in (\ref{scal-gen}), 
while the simple finite-size scaling  form remains valid. 
Our numerical data suggest that this is  the case with  Barkhausen 
avalanches for the range of concentrations in the interior region
between lines ({\it i}) and ({\it ii}) (see Fig.\ 3). We find that 
the following scaling form is satisfied \cite{SOC}
\begin{equation}
D(s,c,L) = L^{-\alpha (c)}{\cal{D}}(sL^{-D_s(c)}) \; .\label{scal-el}
\end{equation}
\noindent
By fitting the data for $L=$ 100, 200, and 300 to the expression 
(\ref{scal-el}) 
we determine the exponent $\alpha (c)$ and the fractal dimension of
avalanches $D_s(c)$ for several values of $c$ (and $f$) in the region II. 
The results are also shown in the inset 
to Fig.\ 1  as a function of $c$ for fixed $f=0.5$. 
For each  value of $f$ a distinct family  of curves is found. 
For instance, the finite-size scaling fit which is shown in Fig.\ 4
is obtained at the point $c=$0.1, $f=$1 (marked by $\star $ on the 
phase diagram), leading  to the exponents 1+$\alpha =$ 
1.54 $\pm $0.02 and $D_s = $1.98 $\pm $0.02, whereas at the same point
we find $\tau  = $1.294 $\pm $  0.003. By comparing the scaling fits for
different $c$ further $c$-dependence of the scaling function in Eq.\ 
(\ref{scal-el})  of the form ${\cal{D}}(x) \sim x^{-\phi (c)}$ for 
$x\to 0$ was found, with numerical values of $\phi (c)$ close to $\tau (c)$.

The scaling region is further characterized by the following scaling 
relation \cite{SC}
\begin{equation}
\alpha (c) = \tau (c)\ D_s(c) \; , \label{scal-eq}
\end{equation}
\noindent
which follows from (\ref{scal-el})  by choosing $sL^{-D_s} \approx 1$ 
and comparing to (\ref{soc_s}). 
It should be stressed that  numerical values of the exponents obtained 
above by the power-law and finite-size scaling fits satisfy (within 
numerical error) the scaling relation  (\ref{scal-eq}) for several 
distinct values of $c$ in the region II (cf. inset to Fig.\ 1).

In the critical region close to the transition line $c_0(f)$ (line 
{\it (i)} in Fig.\ 3) the finite-size scaling form (\ref{scal-el}) fails.
 Instead, a more general  expression (\ref{scal-gen})
with   $c$-dependent coherence length is more appropriate, similar to
the analysis in Refs.\ \cite{num2,num5}. According to our results, the 
width $\delta c$ of the critical region in the direction of the $c$-axis 
is  rather small $\delta c \mathopen< 0.05$, in contrast to the $c=0$ case 
along $f$-axis,  where the fits to the simple scaling form (\ref{scal-el}) 
are not satisfactory in a wide range of $f$. 
This indicates that the scaling and nonuniversality in the region II are 
closely related to the  structure of the underlying spin clusters 
for $c\neq 0$  \cite{comment-asp}.

The scaling region defined via Eqs.\ (\ref{scal-el}) and (\ref{scal-eq})
is also bounded from  the side of strong disorder by the line  
$c^\star (f)$ (line {\it (ii)} in Fig.\ 3),
which ends on the $c$-axis at the percolation threshold $c^\star (0) =0.395$.
Above this line  the power-law is  practically lost  
and strong disorder prevents formation of  system-size avalanches
(cf. the two lower curves in Fig.\ 2).
We expect that in this region the self-similarity persists only on  a 
finite  scale $\xi _s(c)\ll L$, with 
\begin{equation}
D(s,c,L) = s^{-\tau ^\star }{\cal{D}}(s/\xi _s(c),sL ^{-D_s^\star})\; ,
 \label{scal-disordered}
\end{equation}
\noindent
similar to a subcritical cellular automaton \cite{TRmore}.
Here the exponents $\tau ^\star $ and $D_s^\star $ are referring to the 
transition point $c^\star (f)$, and the $c$-dependent correlation length
is expected to behave as $\xi _s(c-c^\star )\sim 
(c-c^\star )^{-D_s^\star }$ \cite{TRmore}. More detailed analysis of 
both critical regions as well as the precise location of the transition 
lines $c_0(f)$ and $c^\star (f)$ is left for a future study.

By varying  the driving rate $\delta h $ at fixed disorder the distributions
$D(s,\delta h)$ appear to have different slopes. For larger $\delta h$ we find  
smaller exponents, for instance, at $c=$0.2, $f=0.5$ the exponent 
1+$\tau  =$1.2904 $\pm $ 0.0015 found with  the driving rate
$\delta h =$0.01,  becomes   $1.2575 \pm  0.0017$ for $\delta h= 0.015$,
and $ 1.239 \pm 0.002$ for $\delta h = 0.02$. Larger   $\delta h$
overdrives weaker pinning centers thus rendering the occurrence of
larger avalanches  more probable.
This picture is in  agreement with the experimental results 
on 1.8\% Si-Fe alloy \cite{ex7Durin}.

We have demonstrated that Barkhausen noise  in {\it diluted} ferromagnetic 
samples at low temperatures  depends on a variety of physical parameters 
which can be  controlled  experimentally, such as strength  of disorder 
and the applied driving rate. The comparison with experiments on 
1.8\% Si-Fe alloy \cite{ex7Durin} where the driving rate was varied, 
is promising. Our numerical data suggest that a more complex behavior 
emerges when the degree of disorder is varied. This includes a nonequilibrium 
phase transition at critical disorder $c_0(f)$ and a true scaling region 
with nonuniversal exponents. Further studies, both theoretical and experimental, 
will help elucidate the general question of conditions that result in the 
breakdown of universality.

I would like to thank Professor Sava Milo\v sevi\'c  for 
stimulating discussions. This work was supported in part by the Ministry 
of Science and Technology of the Republic of Slovenia and by bilateral  
Slovene-German Project No. 1B0A6C.

\end{multicols}

\twocolumn
\begin{figure}\epsfxsize=64mm
\epsffile[40 84 573 725]{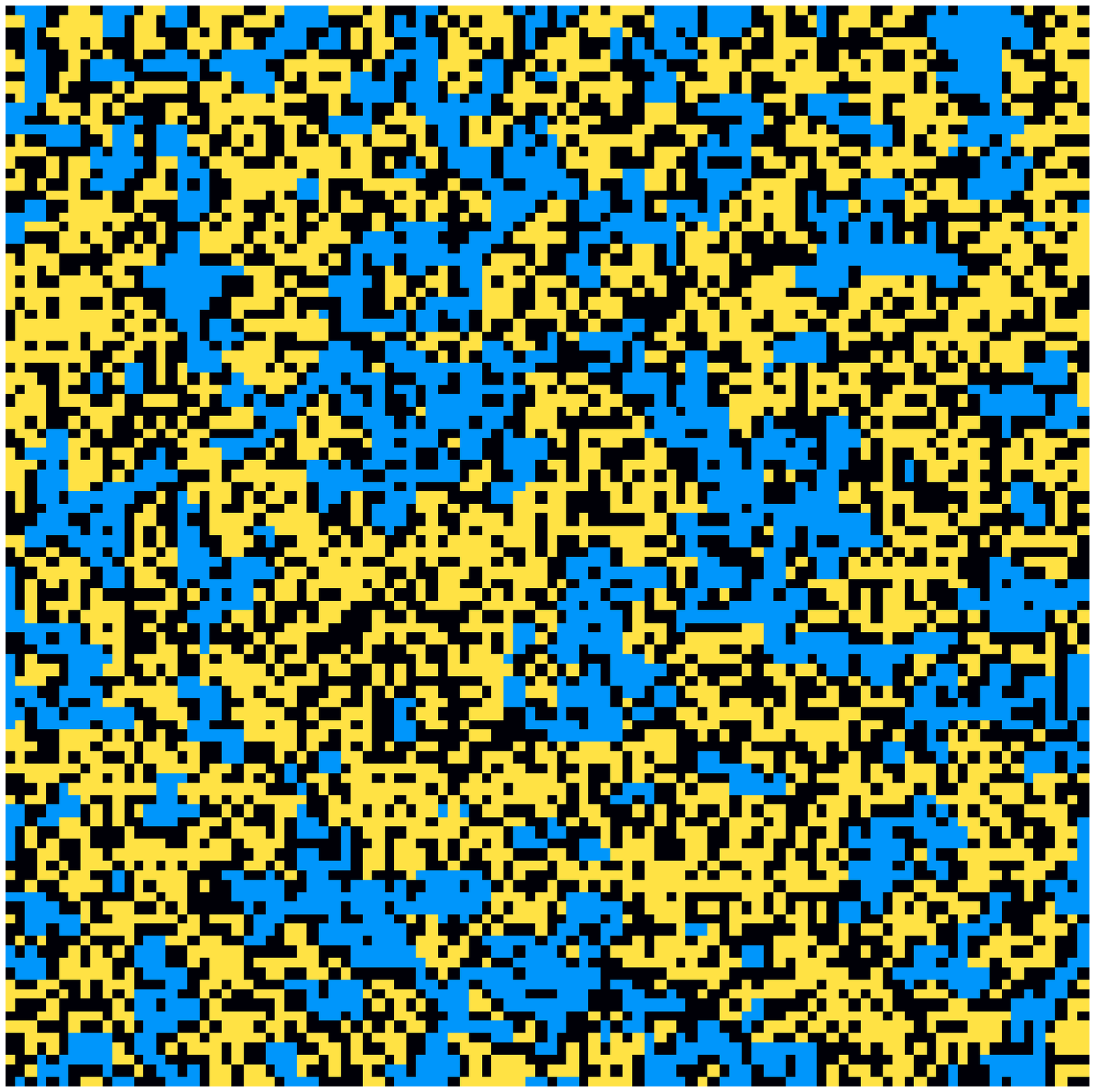}
\vskip 8mm
\caption{\label{fig1}Fractal  cluster of connected spins close to the 
percolation threshold with areas of up spins (bright) and down spins (gray). 
Nonmagnetic defects are shown  as dark  points.}
\end{figure}

\begin{figure}
\epsfxsize=64mm
\epsffile[40 84 573 725]{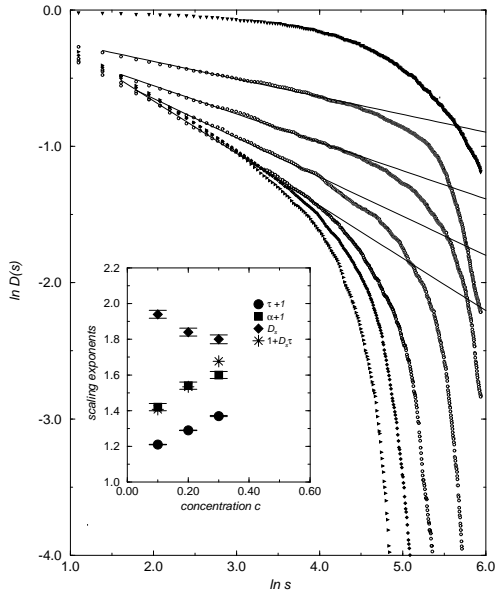}
\vskip 8mm
\caption{\label{fig2}Double logarithmic plot of the integrated distribution 
of size $D(s,c)$ vs. size $s$ of avalanches for fixed lattice size $L=$100 
and $f=0.5$ and for various values of concentration $c$= 0.0, 0.05, 0.1, 
0.2, 0.3, 0.35, and 0.4 (from top to bottom), normalized to the first point. 
Inset: Exponents of distributions of size $1+\tau $, and length $1+\alpha $,
and fractal dimension $D_s$ of avalanches plotted vs. $c$ for fixed $f$=0.5 
in the scaling region.  Also plotted is $1+D_s \ \tau $. }
\end{figure}

\begin{figure}\epsfxsize=64mm
\epsffile[164 278 440 516]{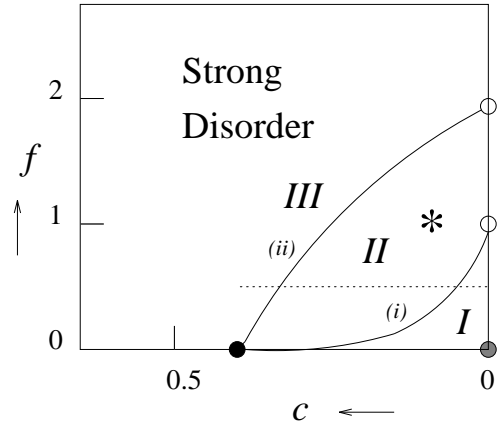}
\vskip 8mm
\caption{\label{fig3} Schematic phase diagram showing (I) region
of  1st order phase transition, (II) scaling region, and (III)
region of strong disorder. The system ceases to percolate along the
line ({\it ii}), while a nonequilibrium phase transition in the
universality class of RFIM occurs along the line ({\it i}). Results 
of Fig\ 2 are obtained by varying concentration $c$ along
the dashed line. The distributions in Fig.\ 4 are  calculated at the 
point marked by $\star $.}
\end{figure}

\begin{figure}\epsfxsize=64mm
\epsffile[40 84 573 725]{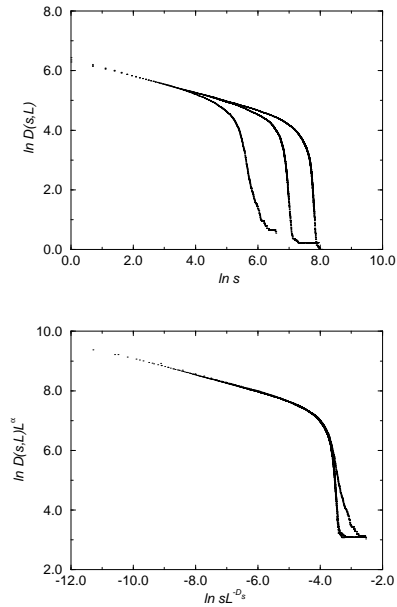}
\vskip 8mm
\caption{\label{fig4}Double logarithmic plot of the integrated 
distribution of size $D(s,c,L)$ vs. $s$ for fixed $c=$0.1 and $f=$1.0 
and three different values of lattice sizes $L=$100, 200, and 300 (top) 
and its finite-size scaling plot according to Eq.\ (\ref{scal-el}) 
(bottom).}
\end{figure}


\begin{references}
\bibitem[18]{thexp}Notice that our $1+\tau (c)$ corresponds to measured 
exponent $\tau _{exp}$ of the probability distribution of sizes.
\bibitem[21]{SC} This scaling relation (among others) was derived in
Ref.\ \cite{ex3Bgg} by assuming the elementary signal of the form
$m(t)=t^{\gamma -1}g(t/T)$. It should be noticed that our exponents are 
defined for the integrated distributions.
\end{references}
\end{document}